\documentclass[acmtog]{acmart}
\acmSubmissionID{1234}

\usepackage{booktabs} 
\graphicspath{ {images/} }
\usepackage{listings}
\usepackage{array}
\usepackage{float}
\usepackage[export]{adjustbox}
\usepackage{indentfirst}
\usepackage{latexsym}
\usepackage{times}
\usepackage{epsfig}
\usepackage{epstopdf}
\usepackage{gensymb}
\usepackage{xcolor}
\usepackage{marvosym}
\usepackage{color}
\usepackage{algorithmic}
\usepackage{graphicx}
\usepackage{textcomp}
\usepackage{xcolor}
\usepackage[utf8]{inputenc} 
\usepackage[T1]{fontenc}    
\usepackage{hyperref}       
\usepackage{url}            
\usepackage{microtype} 

\definecolor{dkgreen}{rgb}{0,0.6,0}
\definecolor{gray}{rgb}{0.5,0.5,0.5}
\definecolor{mauve}{rgb}{0.58,0,0.82}

 \hypersetup{
   colorlinks=false,
   pdfborder={0 0 0},
}

\usepackage[ruled]{algorithm2e} 

\SetAlFnt{\small}
\SetAlCapFnt{\small}
\SetAlCapNameFnt{\small}
\SetAlCapHSkip{0pt}

\acmJournal{TOG}

\begin{document}
\title{Mul2MAR: A Multi-Marker Mobile Augmented Reality Application for Improved Visual Perception}

\author{Murat Kurt}
\orcid{0000-0002-3236-5595}
\affiliation{%
  \institution{International Computer Institute, Ege University}
  \streetaddress{Bornova}
  \city{Izmir}
  \postcode{35100}
  \country{Turkey}}
\email{murat.kurt@ege.edu.tr}

\renewcommand\shortauthors{Kurt, M.}

\begin{abstract}
This paper presents an inexpensive Augmented Reality (AR) application which is aimed to use with mobile devices. Our application is a marker based AR application, and it can be used by inexpensive three dimensional (3D) red-cyan glasses. In our AR application, we combine left and right views without creating any uncomfortable situation for human eyes. We validate our mobile AR application on several objects, scenes, and views. We show that 3D AR perception can be obtained by using our inexpensive AR application~\cite{Gungor2014SIU}.     
\end{abstract}

%
%

\ccsdesc[500]{Computing methodologies~Computer graphics}
\ccsdesc[300]{Computing methodologies~Computer vision}
\ccsdesc[500]{Computing methodologies~Mixed / augmented reality}
\ccsdesc[500]{Computing methodologies~Perception}

%
%

\keywords{Augmented Reality, AR, visual perception, three dimensional visualization}

\maketitle

\section{Introduction}
The use of Augmented Reality (AR) technology has increased considerably in recent years with development of devices such as smart phones, tablets, and smart glasses. The term AR can be defined in many ways, but general meaning of AR is that integrating 3D virtual objects into a 3D physical environment in a real time. AR allows us to  enhance the richness of the real world by using computers~\cite{azuma1997survey,Huma2015}. AR is different from the Virtual Reality (VR). The main difference is that the real world objects are enhanced with the help of virtually in AR, whereas VR offers a digital recreation of a real life and it replaces the real world with a digitally simulated one. AR has begun to be used in many fields with technological developments in recent years. Some of usage fields are military, health sector, industry, manufacturing, museums, entertainment and gaming. AR provides many convenience to user in all used fields and in the upcoming days, use of AR will increase much~\cite{Huma2015}. 

Nowadays, 3D televisions create illusion of three dimensions with using polarized glasses. There are two type of polarized glasses that are active and passive glasses. Before using of these polarized glasses, anaglyph 3D glasses were used. One of this kind of glasses is a red-cyan glass. The red-cyan glass is the most common, inexpensive and widely used than others. Polarized glasses create more effective 3D images than anaglyph 3D glasses. However, they are not suitable for using with mobile devices. They are also very expensive. Unlike them, anaglyph 3D glasses are suitable for use with mobile devices and it is also inexpensive~\cite{Gungor2014SIU}. Many companies produce special AR glasses for different purpose. Some of these are Vuzix M100~\cite{Vuzix}, Meta~\cite{Meta}, Google Glass~\cite{Google} and Apple Vision Pro~\cite{AppleVisionPro}. However, prices of these AR glasses are very high and it is difficult to afford the charge of these glasses for all consumers. 

In this paper, we present an inexpensive AR application which can be used by red-cyan glasses. We prefer to use red-cyan glasses, because it is inexpensive and widely used. We validate our mobile AR application on several objects, scenes, and views by using inexpensive red-cyan glasses. We show that our mobile AR application provides satisfactory 3D AR perception without causing any uncomfortable situation for human eyes.

\section{Related Works}
AR intends to place virtual objects into physical world visuals. To achieve this effect, a software that uses virtual reality combines real world elements with virtual objects on real-time images~\cite{Cawood2008}. Azuma~\cite{azuma1997survey} conducted a poll on his work, and defined three attributes that an ideal AR system should have: combination of reality and virtuality, real-time interactivity and 3D registration.

AR is used on many different fields; to create models and visualize historical buildings in architecture, to visualize human body in medicine, to enhance education materials in engineering, and also in entertainment and games~\cite{azuma1997survey,Broschart2014}. Although this technology seems recent, the first prototype accepted as a VR and AR system was developed by Sutherland~\cite{Sutherland1968}. However, this system was quite expensive and even heavy, needed to be hung on ceiling. Nowadays, AR is developing rapidly thanks to the widespread use of personal computers. In 2009, Christian Doppler Laboratory listed the most important milestones in AR technologies. Even though high-quality applications are being developed, there are not many differences between them and the ones listed~\cite{Doppler}.
\begin{figure*}[t]
  \centering
  \includegraphics[width=0.95\linewidth]{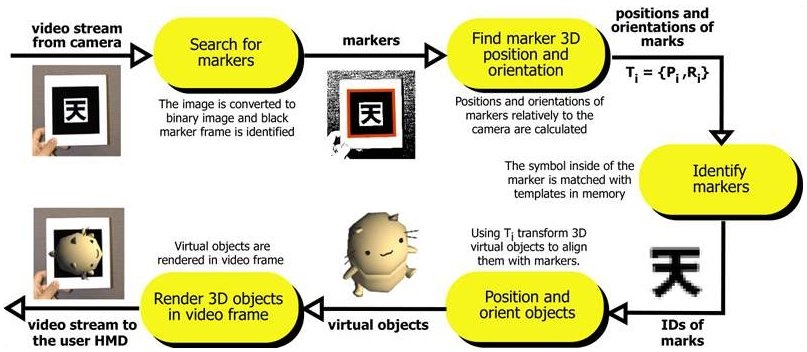}
  \caption{ \label{fig:arsystem}
          A general overview of a marker-based AR system~\cite{ARToolKit}.
           }
\end{figure*}

\section{Method}
In this project we used ARToolKit for our augmented reality technique and OpenGL libraries for the graphics. In Augmented Reality, there are different methods to recognize the environment and place the object. The easiest and cheapest one is to use markers. A general overview of a marker-based AR system can be found in Figure~\ref{fig:arsystem}. Some of the recent works does not even use markers; instead they use Dense Tracking and Mapping (DTAM), but those are more expensive and computationally heavy. In this work, we prefer to use markers.
Markers are important elements in AR. They define where the object will be placed. Usually they are black-and-white geometric patterns, but coloured ones can be used as well. They should be rotationally asymmetric and unique.
Markers represent coordination system for the associated virtual objects in real-time images. When the camera captures an image, it is converted to a black and white image. In this image potential markers are searched by the AR software. If a marker is found, and it matches with the known patterns, software places coordination axes and origin in the image. Lastly, object is placed in the image accordingly~\cite{ARToolKit}. Best results are observed by using contrast and abstract black-and-white markers, since the software we use converts the captured image to black and white image. Coloured markers  may look fuzzy and blend into background. To make the distinction between background and markers, we preferred to use QR codes. As a result, the markers does not get mistaken for background objects, and also does not get mistaken for one another. Compared to traditional markers (see Figure~\ref{fig:arsystem}) using QR codes gives a better result on that case. 

\begin{figure}[t]
  \centering
  \includegraphics[width=0.85\linewidth]{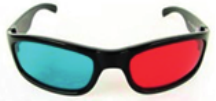}
  \caption{ \label{fig:redcyangl}
          A red-cyan glasses.
           }
\end{figure}

People see with both eyes, but they will not see two different images. Brain form an overlap with images taken from two different angles and it creates the perception of depth. Images taken from two different channels with slightly different angles (like our eyes) can be combined to create 3D effect. Using red-cyan glasses (see Figure~\ref{fig:redcyangl}), right eye sees through cyan filter and only cyan shades (green and blue) passes through it. Similar rule applies for left eye and red filter, which allows only red, orange and yellow pass through. Red and cyan parts of the picture are placed slightly separated from each other, and our brain combines those two different images to achieve the desired 3D effect.

\begin{figure}[t]
  \centering
  \includegraphics[width=0.85\linewidth]{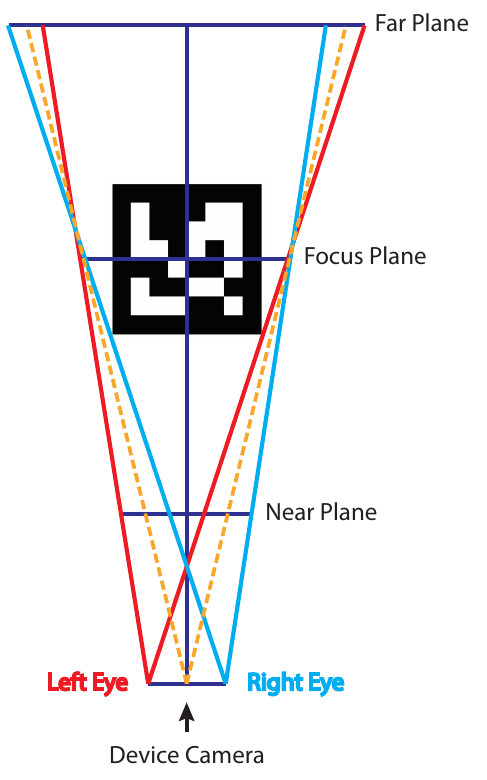}
  \caption{ \label{fig:leftrightview}
          Taking images of 3D objects for the left and the right eyes.
           }
\end{figure}

AR application inserts 3D objects into the center of the marker pattern after calculating its distance and angle to the camera. In our AR method, images are obtained for the left eye and the right eye by OpenGL camera movements without moving device camera, which can be seen in Figure~\ref{fig:leftrightview}.

The following method is used for detection of the marker pattern with device camera in ARToolKit library:
\begin{lstlisting}
arGetTransMat(&marker_info[k], patt_center, patt_width1, patt_trans);
\end{lstlisting}
Here, $\small\texttt{patt\_trans}$ gives the location and angle of the pattern and it's a matrix of $3 \times 4$ dimensions. The following method is used for using of this matrix in OpenGL:
\begin{lstlisting}
argConvGlpara(patt_trans, gl_para);
\end{lstlisting}
Here, $\small\texttt{gl\_para}$ is a vector which has 16 elements. When this vector is transferred to OpenGL modelview matrix with the following method:
\begin{lstlisting}
glLoadMatrixd(gl_para);
\end{lstlisting}
the angle and the distance of the object are set. When an object is placed to the center of the marker pattern, an AR display is shown. However, left and right eyes see different perspectives of the object in a stereo view. To be able to create the stereo view effect, camera is shifted to the left and right eye points by modifying  $\small\texttt{gl\_para}$ vector before it's called by  $\small\texttt{glLoadMatrixd}$ function, as it's seen in Figure~\ref{fig:leftrightview}. 

In order to avoid eyes to see the images of each other, the OpenGL filter methods are applied. For the left eye:
\begin{lstlisting}
glColorMask(true, false, false, false);
\end{lstlisting}
and for the right eye:
\begin{lstlisting}
glColorMask(false, true, true, false);
\end{lstlisting}
In our AR application, when the scene is viewed without red-cyan glasses (see Figure~\ref{fig:redcyangl}), the color sliding appears. When the scene is viewed with red-cyan glasses (see Figure~\ref{fig:redcyangl}), the visual perception of three-dimensional AR is provided~\cite{Gungor2014SIU}.

\section{Results}
\begin{figure*}[t]
  \centering
  \setlength{\tabcolsep}{1pt}
  \setlength{\fboxsep}{0pt}
  \begin{tabular}{cccc}
  \includegraphics[width=.224\linewidth]{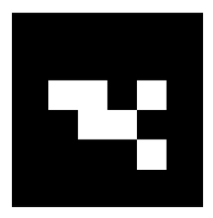}
  &\includegraphics[width=.376\linewidth]{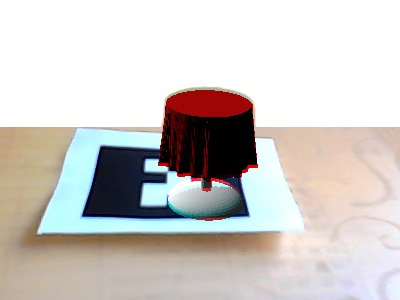}
  &\includegraphics[width=.375\linewidth]{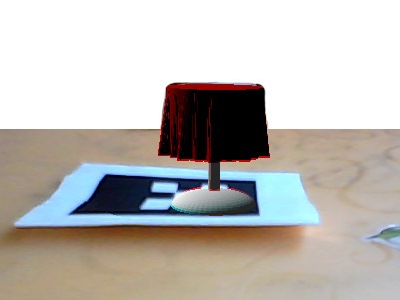}\\
  \mbox{(a)}
  &\mbox{(b)}
  &\mbox{(c)}\\
  \includegraphics[width=.224\linewidth]{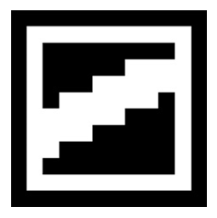}
  &\includegraphics[width=.375\linewidth]{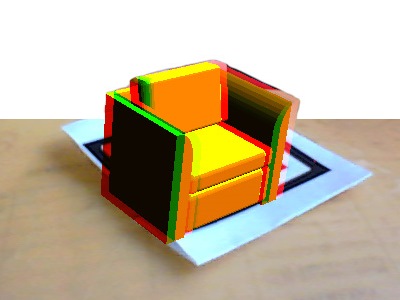}
  &\includegraphics[width=.375\linewidth]{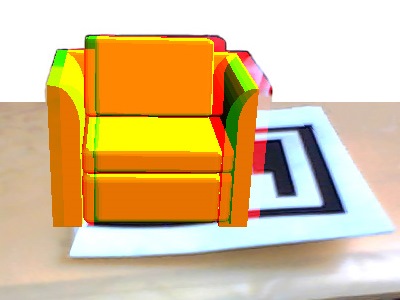}\\
  \mbox{(d)}
  &\mbox{(e)}
  &\mbox{(f)}\\
  \includegraphics[width=.224\linewidth]{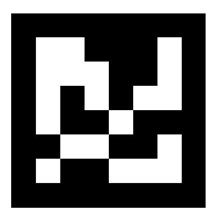}
  &\includegraphics[width=.375\linewidth]{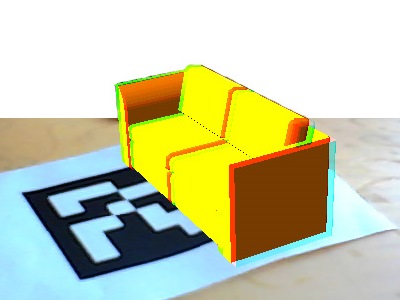}
  &\includegraphics[width=.375\linewidth]{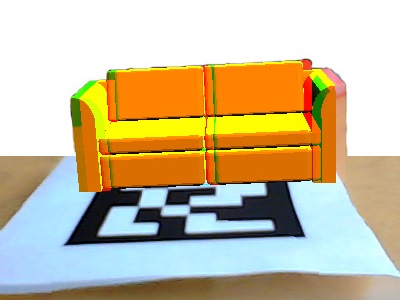}\\
  \mbox{(g)}
  &\mbox{(h)}
  &\mbox{(k)}
  \end{tabular}
  \caption{\label{fig:arobjects}%
           (b), (c), (e), (f), (h), and (k) were rendered by using our marker-based AR application. (a) were used to render (b) and (c). (d) were used to render (e) and (f). (g) were used to render (h) and (k). (b) and (c) are from a table object. (e) and (f) are from a single seat object. (h) and (k) are from a double seat object. We also assigned some material properties and textures to the objects by using OpenGL library. 
           }
\end{figure*}

\begin{figure*}[t]
  \centering
  \includegraphics[width=0.95\linewidth]{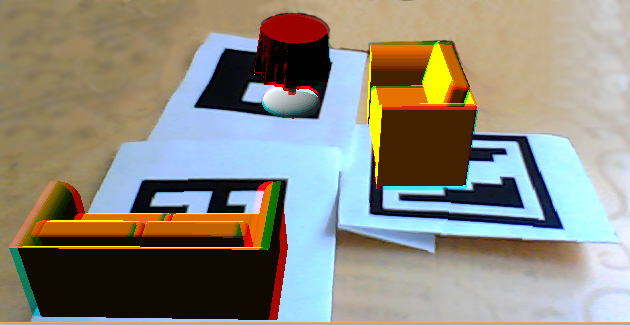}
  \caption{ \label{fig:resulttriple}
          Multiple objects can be rendered by our mobile AR technique.
           }
\end{figure*}

AR images of the objects are rendered by using QR-based markers. We tried to introduce many markers for each objects. The system could not fully recognized some markers and objects were not displayed correctly by using these markers. So we chose our markers which are similar to QR codes. These type of markers provide us more healthy displays than others. The markers which we used are shown in Figure~\ref{fig:arobjects}(a, d, g). To be able to get realistic rendering results, we also assigned some materials properties and made texture mappings to the objects by using OpenGL library functions (see Figure~\ref{fig:arobjects} and Figure~\ref{fig:resulttriple}).

Firstly, only one object image is rendered each time, by adjusting marker angle and its distance to camera, as it can be seen in Figure~\ref{fig:arobjects}. Later, other objects are added and different images are rendered; but in this case we had some problems with object positions. To solve this problem we made changes on the code, translating the objects on their positions of markers and scaling them. Thus, we managed to made several combinations of the objects and were able to render realistic 3D images, which can be seen in Figure~\ref{fig:arobjects} and Figure~\ref{fig:resulttriple}.

As a result of all our studies, we could render multiple objects with our mobile AR technique, as it's shown in Figure~\ref{fig:resulttriple}. When we look at the screen without the red-cyan glass, we see some colorimetric shift on the display. However, when we look at the screen with the red-cyan glasses, 3D image perception is provided and colorimetric shift are eliminated on the display.


\section{Conclusions and Future Works}

In this study, we presented Mul2MAR, which is a mobile marker-based AR technique that works with red-cyan glasses. We validated our AR technique by using three different furniture objects. We assigned some material properties and textures to the objects by using OpenGL library functions. We showed that our AR system can render multiple 3D objects visually plausibly. Three different markers and the red-cyan glasses were used for this system and the results obtained are shown in Figure~\ref{fig:arobjects} and Figure~\ref{fig:resulttriple}.

Red-cyan glasses were preferred because they provide inexpensive solution. Even though this type of glasses can not create the most effective 3D images, they are preferred because of wide availability for end users and they are suitable to use with mobile devices. A simple red-cyan masking technique that was used to view objects was implemented, and with a slight adjustment in the application it can be even turned on and off at will.

There are two downsides using red-cyan glasses. First one is, prolonged use of glasses may cause temporary colour impairment on each eye. Second one, people who use prescribed glasses may not be able to use 3D glasses easily.

In the future, we would like to improve our mobile AR technique by investigating fast and accurate QR code algorithms. We believe that a fast and accurate QR code algorithm will make our mobile AR technique be more stable, faster and more accurate. Additionally, there is potential for upgrading and optimizing our Mul2MAR by incorporating state of the art deep learning methods and
techniques~\cite{Gok2023SIU, Azadvatan2024arXiv, Akdogan2024arXiv, Jabbarli2024arXiv}.

To improve visual perception of our Mul2MAR, we are also interested in implementing realistic Bidirectional Reflectance Distribution Function (BRDF)~\cite{Ozturk2006EGUK, Kurt2007MScThesis, Ozturk2008CG, Kurt2008SIGGRAPHCG, Kurt2009SIGGRAPHCG, Kurt2010SIGGRAPHCG, Ozturk2010GraphiCon, Szecsi2010SCCG, Ozturk2010CGF, Bigili2011CGF, Bilgili2012SCCG, Ergun2012SCCG, Toral2014SIU, Tongbuasirilai2017ICCVW, Kurt2019DEU, Tongbuasirilai2020TVC, Akleman2024arXiv}, Bidirectional Scattering Distribution Function (BSDF)~\cite{WKB12, Ward2014MAM, Kurt2014WLRS, Kurt2016SIGGRAPH, Kurt2017MAM, Kurt2018DEU}, Bidirectional Surface Scattering Reflectance Distribution Function (BSSRDF)~\cite{Kurt2013TPCG, Kurt2013EGSR, Kurt2014PhDThesis, Onel2019PL, Kurt2020MAM, Kurt2021TVC, Yildirim2024arXiv} and multi-layered material~\cite{WKB12, Kurt2016SIGGRAPH, Mir2022DEU} models into our multi-marker mobile AR application.

\section*{Acknowledgment}

The author would like to thank Pelin Karag\"{o}zo\u{g}lu, B\"{u}\c{s}ra Temu\c{c}in, Burcu Daylar and Cengiz G\"{u}ng\"{o}r for their valuable comments and proofreading the paper.

%
%
%
%

\bibliographystyle{ACM-Reference-Format}
\bibliography{arXiv25_Mul2MAR_References}


\begin{thebibliography}{47}


\ifx \showCODEN    \undefined \def \showCODEN     #1{\unskip}     \fi
\ifx \showDOI      \undefined \def \showDOI       #1{#1}\fi
\ifx \showISBNx    \undefined \def \showISBNx     #1{\unskip}     \fi
\ifx \showISBNxiii \undefined \def \showISBNxiii  #1{\unskip}     \fi
\ifx \showISSN     \undefined \def \showISSN      #1{\unskip}     \fi
\ifx \showLCCN     \undefined \def \showLCCN      #1{\unskip}     \fi
\ifx \shownote     \undefined \def \shownote      #1{#1}          \fi
\ifx \showarticletitle \undefined \def \showarticletitle #1{#1}   \fi
\ifx \showURL      \undefined \def \showURL       {\relax}        \fi
\providecommand\bibfield[2]{#2}
\providecommand\bibinfo[2]{#2}
\providecommand\natexlab[1]{#1}
\providecommand\showeprint[2][]{arXiv:#2}

\bibitem[{Akdo{\u{g}}an} and {Kurt}(2024)]%
        {Akdogan2024arXiv}
\bibfield{author}{\bibinfo{person}{Adem {Akdo{\u{g}}an}} {and} \bibinfo{person}{Murat {Kurt}}.} \bibinfo{year}{2024}\natexlab{}.
\newblock \showarticletitle{ExTTNet: A Deep Learning Algorithm for Extracting Table Texts from Invoice Images}.
\newblock \bibinfo{journal}{\emph{arXiv preprint arXiv:2402.02246}}, Article \bibinfo{articleno}{arXiv:2402.02246} (\bibinfo{date}{Feb.} \bibinfo{year}{2024}), \bibinfo{numpages}{arXiv:2402.02246}~pages.
\newblock
\urldef\tempurl%
\url{https://doi.org/10.48550/arXiv.2402.02246}
\showDOI{\tempurl}
\showeprint[arxiv]{2402.02246}~[cs.CV]


\bibitem[Akleman et~al\mbox{.}(2024)]%
        {Akleman2024arXiv}
\bibfield{author}{\bibinfo{person}{Ergun Akleman}, \bibinfo{person}{Murat Kurt}, \bibinfo{person}{Derya Akleman}, \bibinfo{person}{Gary Bruins}, \bibinfo{person}{Sitong Deng}, {and} \bibinfo{person}{Meena Subramanian}.} \bibinfo{year}{2024}\natexlab{}.
\newblock \showarticletitle{Hyper-Realist Rendering: A Theoretical Framework}.
\newblock \bibinfo{journal}{\emph{arXiv preprint arXiv:2401.12853}}, Article \bibinfo{articleno}{arXiv:2401.12853} (\bibinfo{date}{Jan.} \bibinfo{year}{2024}), \bibinfo{numpages}{arXiv:2401.12853}~pages.
\newblock
\urldef\tempurl%
\url{https://doi.org/10.48550/arXiv.2401.12853}
\showDOI{\tempurl}
\showeprint[arxiv]{2401.12853}~[cs.GR]


\bibitem[Apple(2024)]%
        {AppleVisionPro}
Apple \bibinfo{year}{2024}\natexlab{}.
\newblock \bibinfo{title}{Apple Vision Pro User Guide}.
\newblock \bibinfo{howpublished}{\url{https://support.apple.com/tr-tr/guide/apple-vision-pro/welcome/visionos}}.
\newblock


\bibitem[ARToolKit(2024)]%
        {ARToolKit}
ARToolKit \bibinfo{year}{2024}\natexlab{}.
\newblock \bibinfo{title}{ARToolKit Augmented Reality}.
\newblock \bibinfo{howpublished}{\url{http://artoolkit.org/documentation/doku.php?id=3_Marker_Training:marker_about}}.
\newblock


\bibitem[{Azadvatan} and {Kurt}(2024)]%
        {Azadvatan2024arXiv}
\bibfield{author}{\bibinfo{person}{Yashar {Azadvatan}} {and} \bibinfo{person}{Murat {Kurt}}.} \bibinfo{year}{2024}\natexlab{}.
\newblock \showarticletitle{MelNet: A Real-Time Deep Learning Algorithm for Object Detection}.
\newblock \bibinfo{journal}{\emph{arXiv preprint arXiv:2401.17972}}, Article \bibinfo{articleno}{arXiv:2401.17972} (\bibinfo{date}{Jan.} \bibinfo{year}{2024}), \bibinfo{numpages}{arXiv:2401.17972}~pages.
\newblock
\urldef\tempurl%
\url{https://doi.org/10.48550/arXiv.2401.17972}
\showDOI{\tempurl}
\showeprint[arxiv]{2401.17972}~[cs.CV]


\bibitem[Azuma(1997)]%
        {azuma1997survey}
\bibfield{author}{\bibinfo{person}{Ronald~T Azuma}.} \bibinfo{year}{1997}\natexlab{}.
\newblock \showarticletitle{A survey of augmented reality}.
\newblock \bibinfo{journal}{\emph{Presence: Teleoperators and virtual environments}} \bibinfo{volume}{6}, \bibinfo{number}{4} (\bibinfo{year}{1997}), \bibinfo{pages}{355--385}.
\newblock


\bibitem[Bilgili et~al\mbox{.}(2011)]%
        {Bigili2011CGF}
\bibfield{author}{\bibinfo{person}{Ahmet Bilgili}, \bibinfo{person}{Ayd{\i}n {\"O}zt{\"u}rk}, {and} \bibinfo{person}{Murat Kurt}.} \bibinfo{year}{2011}\natexlab{}.
\newblock \showarticletitle{A General {BRDF} Representation Based on Tensor Decomposition}.
\newblock \bibinfo{journal}{\emph{Computer Graphics Forum}} \bibinfo{volume}{30}, \bibinfo{number}{8} (\bibinfo{date}{December} \bibinfo{year}{2011}), \bibinfo{pages}{2427--2439}.
\newblock


\bibitem[Bilgili et~al\mbox{.}(2012)]%
        {Bilgili2012SCCG}
\bibfield{author}{\bibinfo{person}{Ahmet Bilgili}, \bibinfo{person}{Ayd{\i}n \"{O}zt\"{u}rk}, {and} \bibinfo{person}{Murat Kurt}.} \bibinfo{year}{2012}\natexlab{}.
\newblock \showarticletitle{Representing BRDF by wavelet transformation of pair-copula constructions}. In \bibinfo{booktitle}{\emph{Proceedings of the 28th Spring Conference on Computer Graphics}} (Budmerice, Slovakia) \emph{(\bibinfo{series}{SCCG '12})}. \bibinfo{publisher}{ACM}, \bibinfo{address}{New York, NY, USA}, \bibinfo{pages}{63--69}.
\newblock
\showISBNx{978-1-4503-1977-5}
\urldef\tempurl%
\url{https://doi.org/10.1145/2448531.2448539}
\showDOI{\tempurl}


\bibitem[Broschart and Zeile(2014)]%
        {Broschart2014}
\bibfield{author}{\bibinfo{person}{Daniel Broschart} {and} \bibinfo{person}{Peter Zeile}.} \bibinfo{year}{2014}\natexlab{}.
\newblock \showarticletitle{ARchitecture-Augmented Reality Techniques and Use Cases in Architecture and Urban Planning}. In \bibinfo{booktitle}{\emph{Proceedings of REAL CORP 2014}} (Vienna, Austria). \bibinfo{pages}{75--82}.
\newblock
\showISBNx{9783950311075}


\bibitem[Cawood and Fiala(2008)]%
        {Cawood2008}
\bibfield{author}{\bibinfo{person}{S. Cawood} {and} \bibinfo{person}{M. Fiala}.} \bibinfo{year}{2008}\natexlab{}.
\newblock \bibinfo{booktitle}{\emph{Augmented Reality: A Practical Guide}}.
\newblock \bibinfo{publisher}{Pragmatic Bookshelf}.
\newblock
\showISBNx{9781934356036}


\bibitem[{Christian Doppler Laboratory}(2014)]%
        {Doppler}
\bibfield{author}{\bibinfo{person}{{Christian Doppler Laboratory}}.} \bibinfo{year}{2014}\natexlab{}.
\newblock \bibinfo{booktitle}{\emph{Handheld Augmented Reality}}.
\newblock
\urldef\tempurl%
\url{http://studierstube.icg.tugraz.at/handheld_ar}
\showURL{%
\tempurl}


\bibitem[Ergun et~al\mbox{.}(2012)]%
        {Ergun2012SCCG}
\bibfield{author}{\bibinfo{person}{Serkan Ergun}, \bibinfo{person}{Murat Kurt}, {and} \bibinfo{person}{Ayd{\i}n \"{O}zt\"{u}rk}.} \bibinfo{year}{2012}\natexlab{}.
\newblock \showarticletitle{Real-time kd-tree based importance sampling of environment maps}. In \bibinfo{booktitle}{\emph{Proceedings of the 28th Spring Conference on Computer Graphics}} (Budmerice, Slovakia) \emph{(\bibinfo{series}{SCCG '12})}. \bibinfo{publisher}{ACM}, \bibinfo{address}{New York, NY, USA}, \bibinfo{pages}{77--84}.
\newblock
\showISBNx{978-1-4503-1977-5}
\urldef\tempurl%
\url{https://doi.org/10.1145/2448531.2448541}
\showDOI{\tempurl}


\bibitem[G{\"{o}}k et~al\mbox{.}(2023)]%
        {Gok2023SIU}
\bibfield{author}{\bibinfo{person}{G{\"{o}}k{\c{c}}e G{\"{o}}k}, \bibinfo{person}{Selin K{\"{u}}{\c{c}}{\"{u}}k}, \bibinfo{person}{Murat Kurt}, {and} \bibinfo{person}{Ergin Tar{\i}}.} \bibinfo{year}{2023}\natexlab{}.
\newblock \showarticletitle{A U-Net Based Segmentation and Classification Approach over Orthophoto Maps of Archaeological Sites}. In \bibinfo{booktitle}{\emph{Proceedings of the IEEE 31st Signal Processing and Communications Applications Conference}} \emph{(\bibinfo{series}{SIU '23})}. \bibinfo{publisher}{IEEE}, \bibinfo{address}{Istanbul, Turkey}, \bibinfo{pages}{1--4}.
\newblock


\bibitem[Google(2024)]%
        {Google}
Google \bibinfo{year}{2024}\natexlab{}.
\newblock \bibinfo{title}{Google glass}.
\newblock \bibinfo{howpublished}{\url{http://www.google.com/glass/start/}}.
\newblock
\newblock
\shownote{Accessed: 2024-02-28}.


\bibitem[G\"{u}ng\"{o}r and Kurt(2014)]%
        {Gungor2014SIU}
\bibfield{author}{\bibinfo{person}{Cengiz G\"{u}ng\"{o}r} {and} \bibinfo{person}{Murat Kurt}.} \bibinfo{year}{2014}\natexlab{}.
\newblock \showarticletitle{Improving Visual Perception of Augmented Reality on Mobile Devices with 3D Red-Cyan Glasses}. In \bibinfo{booktitle}{\emph{Proceedings of the IEEE 22nd Signal Processing and Communications Applications Conference}} \emph{(\bibinfo{series}{SIU '14})}. \bibinfo{publisher}{IEEE}, \bibinfo{address}{Trabzon, Turkey}, \bibinfo{pages}{1706--1709}.
\newblock


\bibitem[{Jabbarl{\i}} and {Kurt}(2024)]%
        {Jabbarli2024arXiv}
\bibfield{author}{\bibinfo{person}{G{\"u}nel {Jabbarl{\i}}} {and} \bibinfo{person}{Murat {Kurt}}.} \bibinfo{year}{2024}\natexlab{}.
\newblock \showarticletitle{LightFFDNets: Lightweight Convolutional Neural Networks for Rapid Facial Forgery Detection}.
\newblock \bibinfo{journal}{\emph{arXiv preprint arXiv:2411.11826}}, Article \bibinfo{articleno}{arXiv:2411.11826} (\bibinfo{date}{Nov.} \bibinfo{year}{2024}), \bibinfo{numpages}{arXiv:2411.11826}~pages.
\newblock
\urldef\tempurl%
\url{https://doi.org/10.48550/arXiv.2411.11826}
\showDOI{\tempurl}
\showeprint[arxiv]{2411.11826}~[cs.CV]


\bibitem[Kurt(2007)]%
        {Kurt2007MScThesis}
\bibfield{author}{\bibinfo{person}{Murat Kurt}.} \bibinfo{year}{2007}\natexlab{}.
\newblock \emph{\bibinfo{title}{A New Illumination Model in Computer Graphics}}.
\newblock \bibinfo{thesistype}{Master's\ thesis}. \bibinfo{school}{International Computer Institute, Ege University}, \bibinfo{address}{Izmir, Turkey}.
\newblock
\newblock
\shownote{140 pages}.


\bibitem[Kurt(2014a)]%
        {Kurt2014PhDThesis}
\bibfield{author}{\bibinfo{person}{Murat Kurt}.} \bibinfo{year}{2014}\natexlab{a}.
\newblock \emph{\bibinfo{title}{An Efficient Model for Subsurface Scattering in Translucent Materials}}.
\newblock \bibinfo{thesistype}{Ph.\,D. Dissertation}. \bibinfo{school}{International Computer Institute, Ege University}, \bibinfo{address}{Izmir, Turkey}.
\newblock
\newblock
\shownote{122 pages}.


\bibitem[Kurt(2014b)]%
        {Kurt2014WLRS}
\bibfield{author}{\bibinfo{person}{Murat Kurt}.} \bibinfo{year}{2014}\natexlab{b}.
\newblock \bibinfo{title}{Grand Challenges in BSDF Measurement and Modeling}.
\newblock \bibinfo{howpublished}{The Workshop on Light Redirection and Scatter: Measurement, Modeling, Simulation}.
\newblock
\newblock
\shownote{(Invited Talk)}.


\bibitem[Kurt(2017)]%
        {Kurt2017MAM}
\bibfield{author}{\bibinfo{person}{Murat Kurt}.} \bibinfo{year}{2017}\natexlab{}.
\newblock \showarticletitle{{Experimental Analysis of BSDF Models}}. In \bibinfo{booktitle}{\emph{Proceedings of the 5th Eurographics Workshop on Material Appearance Modeling: Issues and Acquisition}} \emph{(\bibinfo{series}{MAM '17})}, \bibfield{editor}{\bibinfo{person}{Reinhard Klein} {and} \bibinfo{person}{Holly Rushmeier}} (Eds.). \bibinfo{publisher}{The Eurographics Association}, \bibinfo{address}{Helsinki, Finland}, \bibinfo{pages}{35--39}.
\newblock
\showISBNx{978-3-03868-035-2}
\urldef\tempurl%
\url{https://doi.org/10.2312/mam.20171330}
\showDOI{\tempurl}


\bibitem[Kurt(2018)]%
        {Kurt2018DEU}
\bibfield{author}{\bibinfo{person}{Murat Kurt}.} \bibinfo{year}{2018}\natexlab{}.
\newblock \showarticletitle{A Survey of BSDF Measurements and Representations}.
\newblock \bibinfo{journal}{\emph{Journal of Science and Engineering}} \bibinfo{volume}{20}, \bibinfo{number}{58} (\bibinfo{date}{January} \bibinfo{year}{2018}), \bibinfo{pages}{87--102}.
\newblock
\showISSN{1302-9304}


\bibitem[Kurt(2019)]%
        {Kurt2019DEU}
\bibfield{author}{\bibinfo{person}{Murat Kurt}.} \bibinfo{year}{2019}\natexlab{}.
\newblock \showarticletitle{Real-Time Shading with Phong BRDF Model}.
\newblock \bibinfo{journal}{\emph{Journal of Science and Engineering}} \bibinfo{volume}{21}, \bibinfo{number}{63} (\bibinfo{date}{September} \bibinfo{year}{2019}), \bibinfo{pages}{859--867}.
\newblock
\showISSN{1302-9304}


\bibitem[Kurt(2020)]%
        {Kurt2020MAM}
\bibfield{author}{\bibinfo{person}{Murat Kurt}.} \bibinfo{year}{2020}\natexlab{}.
\newblock \showarticletitle{{A Genetic Algorithm Based Heterogeneous Subsurface Scattering Representation}}. In \bibinfo{booktitle}{\emph{Proceedings of the 8th Eurographics Workshop on Material Appearance Modeling: Issues and Acquisition}} \emph{(\bibinfo{series}{MAM '20})}, \bibfield{editor}{\bibinfo{person}{Reinhard Klein} {and} \bibinfo{person}{Holly Rushmeier}} (Eds.). \bibinfo{publisher}{The Eurographics Association}, \bibinfo{address}{London, UK}, \bibinfo{pages}{13--16}.
\newblock
\showISBNx{978-3-03868-108-3}
\urldef\tempurl%
\url{https://doi.org/10.2312/mam.20201140}
\showDOI{\tempurl}


\bibitem[Kurt(2021)]%
        {Kurt2021TVC}
\bibfield{author}{\bibinfo{person}{Murat Kurt}.} \bibinfo{year}{2021}\natexlab{}.
\newblock \showarticletitle{GenSSS: a genetic algorithm for measured subsurface scattering representation}.
\newblock \bibinfo{journal}{\emph{The Visual Computer}} \bibinfo{volume}{37}, \bibinfo{number}{2} (\bibinfo{date}{February} \bibinfo{year}{2021}), \bibinfo{pages}{307--323}.
\newblock
\urldef\tempurl%
\url{https://doi.org/10.1007/s00371-020-01800-0}
\showDOI{\tempurl}


\bibitem[Kurt and Cinsdikici(2008)]%
        {Kurt2008SIGGRAPHCG}
\bibfield{author}{\bibinfo{person}{Murat Kurt} {and} \bibinfo{person}{Muhammed~G\"{o}khan Cinsdikici}.} \bibinfo{year}{2008}\natexlab{}.
\newblock \showarticletitle{Representing BRDFs using SOMs and MANs}.
\newblock \bibinfo{journal}{\emph{SIGGRAPH Computer Graphics}} \bibinfo{volume}{42}, \bibinfo{number}{3} (\bibinfo{date}{August} \bibinfo{year}{2008}), \bibinfo{pages}{1--18}.
\newblock
\showISSN{0097-8930}
\urldef\tempurl%
\url{https://doi.org/10.1145/1408626.1408630}
\showDOI{\tempurl}


\bibitem[Kurt and Edwards(2009)]%
        {Kurt2009SIGGRAPHCG}
\bibfield{author}{\bibinfo{person}{Murat Kurt} {and} \bibinfo{person}{Dave Edwards}.} \bibinfo{year}{2009}\natexlab{}.
\newblock \showarticletitle{A survey of BRDF models for computer graphics}.
\newblock \bibinfo{journal}{\emph{SIGGRAPH Computer Graphics}} \bibinfo{volume}{43}, \bibinfo{number}{2} (\bibinfo{date}{May} \bibinfo{year}{2009}), \bibinfo{pages}{1--7}.
\newblock
\showISSN{0097-8930}
\urldef\tempurl%
\url{https://doi.org/10.1145/1629216.1629222}
\showDOI{\tempurl}


\bibitem[Kurt and \"{O}zt\"{u}rk(2013)]%
        {Kurt2013EGSR}
\bibfield{author}{\bibinfo{person}{Murat Kurt} {and} \bibinfo{person}{Ayd{\i}n \"{O}zt\"{u}rk}.} \bibinfo{year}{2013}\natexlab{}.
\newblock \showarticletitle{A Heterogeneous Subsurface Scattering Representation Based on Compact and Efficient Matrix Factorization}. In \bibinfo{booktitle}{\emph{Proceedings of the 24th Eurographics Symposium on Rendering, Posters}} \emph{(\bibinfo{series}{EGSR '13})}. \bibinfo{publisher}{Eurographics Association}, \bibinfo{address}{Zaragoza, Spain}.
\newblock


\bibitem[Kurt et~al\mbox{.}(2013)]%
        {Kurt2013TPCG}
\bibfield{author}{\bibinfo{person}{Murat Kurt}, \bibinfo{person}{Ayd{\i}n \"{O}zt\"{u}rk}, {and} \bibinfo{person}{Pieter Peers}.} \bibinfo{year}{2013}\natexlab{}.
\newblock \showarticletitle{A Compact Tucker-Based Factorization Model for Heterogeneous Subsurface Scattering}. In \bibinfo{booktitle}{\emph{Proceedings of the 11th Theory and Practice of Computer Graphics}} \emph{(\bibinfo{series}{TPCG '13})}, \bibfield{editor}{\bibinfo{person}{Silvester Czanner} {and} \bibinfo{person}{Wen Tang}} (Eds.). \bibinfo{publisher}{Eurographics Association}, \bibinfo{address}{Bath, United Kingdom}, \bibinfo{pages}{85--92}.
\newblock
\showISBNx{978-3-905673-98-2}
\urldef\tempurl%
\url{https://doi.org/10.2312/LocalChapterEvents.TPCG.TPCG13.085-092}
\showDOI{\tempurl}


\bibitem[Kurt et~al\mbox{.}(2010)]%
        {Kurt2010SIGGRAPHCG}
\bibfield{author}{\bibinfo{person}{Murat Kurt}, \bibinfo{person}{L\'{a}szl\'{o} Szirmay-Kalos}, {and} \bibinfo{person}{Jaroslav K\v{r}iv\'{a}nek}.} \bibinfo{year}{2010}\natexlab{}.
\newblock \showarticletitle{An Anisotropic BRDF Model for Fitting and Monte Carlo Rendering}.
\newblock \bibinfo{journal}{\emph{SIGGRAPH Computer Graphics}} \bibinfo{volume}{44}, \bibinfo{number}{1} (\bibinfo{date}{February} \bibinfo{year}{2010}), \bibinfo{pages}{1--15}.
\newblock
\showISSN{0097-8930}
\urldef\tempurl%
\url{https://doi.org/10.1145/1722991.1722996}
\showDOI{\tempurl}


\bibitem[Kurt et~al\mbox{.}(2016)]%
        {Kurt2016SIGGRAPH}
\bibfield{author}{\bibinfo{person}{Murat Kurt}, \bibinfo{person}{Greg Ward}, {and} \bibinfo{person}{Nicolas Bonneel}.} \bibinfo{year}{2016}\natexlab{}.
\newblock \showarticletitle{A Data-Driven BSDF Framework}. In \bibinfo{booktitle}{\emph{Proceedings of the ACM SIGGRAPH 2016, Posters}} (Anaheim, California) \emph{(\bibinfo{series}{SIGGRAPH '16})}. \bibinfo{publisher}{ACM}, \bibinfo{address}{New York, NY, USA}, Article \bibinfo{articleno}{31}, \bibinfo{numpages}{2}~pages.
\newblock
\showISBNx{978-1-4503-4371-8}
\urldef\tempurl%
\url{https://doi.org/10.1145/2945078.2945109}
\showDOI{\tempurl}


\bibitem[Meta(2024)]%
        {Meta}
Meta \bibinfo{year}{2024}\natexlab{}.
\newblock \bibinfo{title}{Meta}.
\newblock \bibinfo{howpublished}{\url{https://www.metavision.com//}}.
\newblock
\newblock
\shownote{Accessed: 2024-02-28}.


\bibitem[Mir et~al\mbox{.}(2022)]%
        {Mir2022DEU}
\bibfield{author}{\bibinfo{person}{Sermet Mir}, \bibinfo{person}{Bar{\i}\c{s} Y{\i}ld{\i}r{\i}m}, {and} \bibinfo{person}{Murat Kurt}.} \bibinfo{year}{2022}\natexlab{}.
\newblock \showarticletitle{An Analysis of Goniochromatic and Sparkle Effects on Multi-Layered Materials}.
\newblock \bibinfo{journal}{\emph{Journal of Science and Engineering}} \bibinfo{volume}{24}, \bibinfo{number}{72} (\bibinfo{date}{September} \bibinfo{year}{2022}), \bibinfo{pages}{737--746}.
\newblock
\showISSN{1302-9304}


\bibitem[\"{O}nel et~al\mbox{.}(2019)]%
        {Onel2019PL}
\bibfield{author}{\bibinfo{person}{Sermet \"{O}nel}, \bibinfo{person}{Murat Kurt}, {and} \bibinfo{person}{Ayd{\i}n \"{O}zt\"{u}rk}.} \bibinfo{year}{2019}\natexlab{}.
\newblock \showarticletitle{An Efficient Plugin for Representing Heterogeneous Translucent Materials}.
\newblock In \bibinfo{booktitle}{\emph{Contemporary Topics in Computer Graphics and Games: Selected Papers from the Eurasia Graphics Conference Series}}, \bibfield{editor}{\bibinfo{person}{Veysi \.{I}\c{s}ler}, \bibinfo{person}{Ha\c{s}met G\"{u}r\c{c}ay}, \bibinfo{person}{Hasan~Kemal S\"{u}her}, {and} \bibinfo{person}{G\"{u}ven \c{C}atak}} (Eds.). \bibinfo{publisher}{Peter Lang GmbH, Internationaler Verlag der Wissenschaften}, Chapter~18, \bibinfo{pages}{309--321}.
\newblock
\showISBNx{978-3631802120}
\newblock
\shownote{(Book Chapter)}.


\bibitem[{\"O}zt{\"u}rk et~al\mbox{.}(2006)]%
        {Ozturk2006EGUK}
\bibfield{author}{\bibinfo{person}{Aydin {\"O}zt{\"u}rk}, \bibinfo{person}{Ahmet Bilgili}, {and} \bibinfo{person}{Murat Kurt}.} \bibinfo{year}{2006}\natexlab{}.
\newblock \showarticletitle{Polynomial Approximation of Blinn-Phong Model}. In \bibinfo{booktitle}{\emph{Proceedings of the 4th Theory and Practice of Computer Graphics}} \emph{(\bibinfo{series}{TPCG '06})}, \bibfield{editor}{\bibinfo{person}{Louise~M. Lever} {and} \bibinfo{person}{Mary McDerby}} (Eds.). \bibinfo{publisher}{Eurographics Association}, \bibinfo{address}{Middlesbrough, United Kingdom}, \bibinfo{pages}{55--61}.
\newblock
\showISBNx{3-905673-59-2}
\showISSN{undefined}
\urldef\tempurl%
\url{https://doi.org/10.2312/LocalChapterEvents/TPCG/TPCG06/055-061}
\showDOI{\tempurl}


\bibitem[{\"O}zt{\"u}rk et~al\mbox{.}(2010a)]%
        {Ozturk2010CGF}
\bibfield{author}{\bibinfo{person}{Ayd{\i}n {\"O}zt{\"u}rk}, \bibinfo{person}{Murat Kurt}, {and} \bibinfo{person}{Ahmet Bilgili}.} \bibinfo{year}{2010}\natexlab{a}.
\newblock \showarticletitle{A Copula-Based BRDF Model}.
\newblock \bibinfo{journal}{\emph{Computer Graphics Forum}} \bibinfo{volume}{29}, \bibinfo{number}{6} (\bibinfo{date}{September} \bibinfo{year}{2010}), \bibinfo{pages}{1795--1806}.
\newblock


\bibitem[{\"O}zt{\"u}rk et~al\mbox{.}(2010b)]%
        {Ozturk2010GraphiCon}
\bibfield{author}{\bibinfo{person}{Ayd{\i}n {\"O}zt{\"u}rk}, \bibinfo{person}{Murat Kurt}, {and} \bibinfo{person}{Ahmet Bilgili}.} \bibinfo{year}{2010}\natexlab{b}.
\newblock \showarticletitle{Modeling BRDF by a Probability Distribution}. In \bibinfo{booktitle}{\emph{Proceedings of the 20th International Conference on Computer Graphics and Vision}}. \bibinfo{address}{St. Petersburg, Russia}, \bibinfo{pages}{57--63}.
\newblock


\bibitem[Ozturk et~al\mbox{.}(2008)]%
        {Ozturk2008CG}
\bibfield{author}{\bibinfo{person}{Aydin Ozturk}, \bibinfo{person}{Murat Kurt}, \bibinfo{person}{Ahmet Bilgili}, {and} \bibinfo{person}{Cengiz Gungor}.} \bibinfo{year}{2008}\natexlab{}.
\newblock \showarticletitle{Linear approximation of Bidirectional Reflectance Distribution Functions}.
\newblock \bibinfo{journal}{\emph{Computers \& Graphics}} \bibinfo{volume}{32}, \bibinfo{number}{2} (\bibinfo{date}{April} \bibinfo{year}{2008}), \bibinfo{pages}{149--158}.
\newblock


\bibitem[Shoaib and Jaffry(2015)]%
        {Huma2015}
\bibfield{author}{\bibinfo{person}{Huma Shoaib} {and} \bibinfo{person}{S.~Waqar Jaffry}.} \bibinfo{year}{2015}\natexlab{}.
\newblock \showarticletitle{A Survey of Augmented Reality}. In \bibinfo{booktitle}{\emph{Proceedings of the International Conference on Virtual and Augmented Reality}} \emph{(\bibinfo{series}{ICVAR 2015})}. \bibinfo{publisher}{IEEE}, \bibinfo{address}{Singapore}, \bibinfo{pages}{1706--1709}.
\newblock


\bibitem[Sutherland(1968)]%
        {Sutherland1968}
\bibfield{author}{\bibinfo{person}{I.~E. Sutherland}.} \bibinfo{year}{1968}\natexlab{}.
\newblock \showarticletitle{A Head-Mounted Three Dimensional Display}. In \bibinfo{booktitle}{\emph{Proceedings of the Fall Joint Computer Conference}} (San Francisco, California) \emph{(\bibinfo{series}{AFIPS '68 (Fall, part I)})}. \bibinfo{publisher}{ACM}, \bibinfo{address}{New York, NY, USA}, \bibinfo{pages}{757--764}.
\newblock
\urldef\tempurl%
\url{https://doi.org/10.1145/1476589.1476686}
\showDOI{\tempurl}


\bibitem[Sz\'{e}csi et~al\mbox{.}(2010)]%
        {Szecsi2010SCCG}
\bibfield{author}{\bibinfo{person}{L\'{a}szl\'{o} Sz\'{e}csi}, \bibinfo{person}{L\'{a}szl\'{o} Szirmay-Kalos}, \bibinfo{person}{Murat Kurt}, {and} \bibinfo{person}{Bal\'{a}zs Cs\'{e}bfalvi}.} \bibinfo{year}{2010}\natexlab{}.
\newblock \showarticletitle{Adaptive sampling for environment mapping}. In \bibinfo{booktitle}{\emph{Proceedings of the 26th Spring Conference on Computer Graphics}} (Budmerice, Slovakia) \emph{(\bibinfo{series}{SCCG '10})}. \bibinfo{publisher}{ACM}, \bibinfo{address}{New York, NY, USA}, \bibinfo{pages}{69--76}.
\newblock
\showISBNx{978-1-4503-0558-7}
\urldef\tempurl%
\url{https://doi.org/10.1145/1925059.1925073}
\showDOI{\tempurl}


\bibitem[Tongbuasirilai et~al\mbox{.}(2020)]%
        {Tongbuasirilai2020TVC}
\bibfield{author}{\bibinfo{person}{Tanaboon Tongbuasirilai}, \bibinfo{person}{Jonas Unger}, \bibinfo{person}{Joel Kronander}, {and} \bibinfo{person}{Murat Kurt}.} \bibinfo{year}{2020}\natexlab{}.
\newblock \showarticletitle{Compact and intuitive data-driven BRDF models}.
\newblock \bibinfo{journal}{\emph{The Visual Computer}} \bibinfo{volume}{36}, \bibinfo{number}{4} (\bibinfo{date}{April} \bibinfo{year}{2020}), \bibinfo{pages}{855--872}.
\newblock
\urldef\tempurl%
\url{https://doi.org/10.1007/s00371-019-01664-z}
\showDOI{\tempurl}


\bibitem[Tongbuasirilai et~al\mbox{.}(2017)]%
        {Tongbuasirilai2017ICCVW}
\bibfield{author}{\bibinfo{person}{Tanaboon Tongbuasirilai}, \bibinfo{person}{Jonas Unger}, {and} \bibinfo{person}{Murat Kurt}.} \bibinfo{year}{2017}\natexlab{}.
\newblock \showarticletitle{Efficient {BRDF} Sampling Using Projected Deviation Vector Parameterization}. In \bibinfo{booktitle}{\emph{Proceedings of the {IEEE} International Conference on Computer Vision Workshops}} \emph{(\bibinfo{series}{ICCVW '17})}. \bibinfo{publisher}{{IEEE} Computer Society}, \bibinfo{address}{Venice, Italy}, \bibinfo{pages}{153--158}.
\newblock
\showISBNx{978-1-5386-1034-3}
\urldef\tempurl%
\url{https://doi.org/10.1109/ICCVW.2017.26}
\showDOI{\tempurl}


\bibitem[T\"{o}ral et~al\mbox{.}(2014)]%
        {Toral2014SIU}
\bibfield{author}{\bibinfo{person}{\"{O}zkan~An{\i}l T\"{o}ral}, \bibinfo{person}{Serkan Ergun}, \bibinfo{person}{Murat Kurt}, {and} \bibinfo{person}{Ayd{\i}n \"{O}zt\"{u}rk}.} \bibinfo{year}{2014}\natexlab{}.
\newblock \showarticletitle{Mobile GPU-Based Importance Sampling}. In \bibinfo{booktitle}{\emph{Proceedings of the IEEE 22nd Signal Processing and Communications Applications Conference}} \emph{(\bibinfo{series}{SIU '14})}. \bibinfo{publisher}{IEEE}, \bibinfo{address}{Trabzon, Turkey}, \bibinfo{pages}{510--513}.
\newblock


\bibitem[Vuzix(2024)]%
        {Vuzix}
Vuzix \bibinfo{year}{2024}\natexlab{}.
\newblock \bibinfo{title}{Vuzix}.
\newblock \bibinfo{howpublished}{\url{https://www.vuzix.com/}}.
\newblock
\newblock
\shownote{Accessed: 2024-02-28}.


\bibitem[Ward et~al\mbox{.}(2012)]%
        {WKB12}
\bibfield{author}{\bibinfo{person}{Greg Ward}, \bibinfo{person}{Murat Kurt}, {and} \bibinfo{person}{Nicolas Bonneel}.} \bibinfo{year}{2012}\natexlab{}.
\newblock \bibinfo{booktitle}{\emph{A Practical Framework for Sharing and Rendering Real-World Bidirectional Scattering Distribution Functions}}.
\newblock \bibinfo{type}{{T}echnical {R}eport} LBNL-5954E. \bibinfo{institution}{Lawrence Berkeley National Laboratory}.
\newblock


\bibitem[Ward et~al\mbox{.}(2014)]%
        {Ward2014MAM}
\bibfield{author}{\bibinfo{person}{Greg Ward}, \bibinfo{person}{Murat Kurt}, {and} \bibinfo{person}{Nicolas Bonneel}.} \bibinfo{year}{2014}\natexlab{}.
\newblock \showarticletitle{Reducing Anisotropic BSDF Measurement to Common Practice}. In \bibinfo{booktitle}{\emph{Proceedings of the 2nd Eurographics Workshop on Material Appearance Modeling: Issues and Acquisition}} \emph{(\bibinfo{series}{MAM '14})}, \bibfield{editor}{\bibinfo{person}{Reinhard Klein} {and} \bibinfo{person}{Holly Rushmeier}} (Eds.). \bibinfo{publisher}{Eurographics Association}, \bibinfo{address}{Lyon, France}, \bibinfo{pages}{5--8}.
\newblock
\showISBNx{978-3-905674-64-4}
\urldef\tempurl%
\url{https://doi.org/10.2312/mam.20141292}
\showDOI{\tempurl}


\bibitem[{Y{\i}ld{\i}r{\i}m} and {Kurt}(2024)]%
        {Yildirim2024arXiv}
\bibfield{author}{\bibinfo{person}{Bar{\i}{\c{s}} {Y{\i}ld{\i}r{\i}m}} {and} \bibinfo{person}{Murat {Kurt}}.} \bibinfo{year}{2024}\natexlab{}.
\newblock \showarticletitle{GenPluSSS: A Genetic Algorithm Based Plugin for Measured Subsurface Scattering Representation}.
\newblock \bibinfo{journal}{\emph{arXiv preprint arXiv:2401.15245}}, Article \bibinfo{articleno}{arXiv:2401.15245} (\bibinfo{date}{Jan.} \bibinfo{year}{2024}), \bibinfo{numpages}{arXiv:2401.15245}~pages.
\newblock
\urldef\tempurl%
\url{https://doi.org/10.48550/arXiv.2401.15245}
\showDOI{\tempurl}
\showeprint[arxiv]{2401.15245}~[cs.GR]


\end{thebibliography}

\end{document}